\documentclass[10pt, prb, aps, twocolumn, showpacs, citeautoscript, floatfix, reprint, amsmath, amssymb, notitlepage, superscriptaddress, longbibliography]{revtex4-1}

\usepackage{graphicx}
\usepackage{stmaryrd}
\usepackage{rotating}
\usepackage{amsmath}
\usepackage{amsfonts}
\usepackage{amssymb}
\usepackage{wasysym}
\usepackage[countmax]{subfloat}
\usepackage{dcolumn} 
\usepackage{bm} 
\usepackage{color}

\usepackage{float}
\usepackage{latexsym}

\begin{document}

\title{Real space quantum metric of solids}







\author{Lucas A. Oliveira}

\affiliation{Department of Physics, PUC-Rio, 22451-900 Rio de Janeiro, Brazil}

\author{Wei Chen}

\affiliation{Department of Physics, PUC-Rio, 22451-900 Rio de Janeiro, Brazil}

\date{\rm\today}

\begin{abstract}

By applying the projector to the filled lattice eigenstates on a specific position, or applying the local electron annihilation operator on the many-body ground state, one can construct a quantum state localized around a specific position in a solid. The overlap of two such local states at slightly different positions defines a quantum metric in real space, which manifests even in systems as simple as particles in a box. For continuous systems like electron gas, this metric weighted by the density gives the momentum variance of electrons, which is readily measurable by ARPES. The presence of disorder curves the real space manifold and gives rise to various differential geometrical quantities like Riemann tensor and Ricci scalar, indicating the possibility of engineering differential geometrical properties by disorder, as demonstrated by lattice models of 2D metals and topological insulators.




\end{abstract}

\maketitle

\section{Introduction}

The geometrical properties of the Bloch state $|\psi({\bf k})\rangle$ in the momentum space of solids have been a fascinating subject that connect many seeming unrelated phenomena in condensed matter physics. Starting from a fully gapped semiconductor or insulator and denoting the fully antisymmetric valence band state at momentum ${\bf k}$ by $|\psi({\bf k})\rangle$, the overlap of two such states at slightly different momenta $|\langle\psi({\bf k})|\psi({\bf k+\delta k})\rangle|=1-g_{\mu\nu}\delta k^{\mu}\delta k^{\nu}/2$ (repeated Greek indices are summed) defines the quantum metric $g_{\mu\nu}({\bf k})$ in the momentum space\cite{Provost80}. This quantum metric has been linked to various prominent properties such as topological order\cite{Ma13,Ma14,Yang15,Piechon16,Panahiyan20_fidelity,vonGersdorff21_metric_curvature,Mera22,deSousa23_graphene_opacity,Tan19}, dielectric and optical properties\cite{Ozawa18,Ghosh24,Komissarov24,Chen24_optical_marker}, spread of Wannier function\cite{Marzari97,Marzari12,Souza08}, among many others. Moreover, from this metric, one can proceed to introduce various differential geometrical quantities to the $D$-dimensional momentum space treated as a $T^{D}$ torus, such as the Euler characteristic\cite{Matsuura10,Kolodrubetz13,Kolodrubetz17,Chen25_quantum_geometry_TI_TSC}, allowing the aspects of differential geometry to be investigated entirely in a quantum setting.

Along this line of development, a fundamental question naturally arises: Instead of momentum space, is it possible to define quantum metric in the real space of solids? Shall this be possible, it would have a lot of profound implications. Firstly, a real space quantum metric would be more in line with Einstein's spacetime metric in general relativity that is the origin of countless fascinating phenomena in astrophysics, and also more analogous to the metric on the surface of any 3D object in our daily life. Secondly, the aforementioned momentum space quantum metric is a property of the Bloch state that cannot be easily manipulated, but the real space metric is presumably influenced by defects, much like the deformation of spacetime metric by stars in an otherwise empty space, or poking an elastic object to change the curvature of its surface, opening the possibility to engineering differential geometrical properties by disorder. Motivated by this line of thinking, it has been shown that the momentum space quantum metric integrated over the Brillouin zone can be mapped to real space as a local quantity\cite{Marrazzo19,deSousa23_fidelity_marker}. However, this quantity does not have the physical meaning as the measure of distance between some neighboring quantum states in real space, so one shall seek for some alternative constructions of a metric that can implement Euclidean geometry.

In this paper, we present a local state formalism to introduce a real space quantum metric for solids. We demonstrate that quantum geometry can be introduced directly from a state constructed by acting the projectors to the filled eigenstates on a specific position, or equivalently by acting the electron annihilation operator of a specific position on the ground state. This state is expected to be localized around the specified position, as discussed intensively within the context of density functional theory, theory of charge polarization, and topological markers\cite{Kohn96,Resta06,Thonhauser06,Resta11,Bianco11}, and has the physical meaning of partitioning the ground state expectation value of any quantity (such as energy, momentum, spin, etc.) into each position. This metric manifests ubiquitously in any solids, even in systems as trivial as particles in a box, indicating the ubiquity of our formalism. In continuous metallic systems like electron gases, the metric is found to be determined by the Fermi momentum, and has a remarkable physical interpretation as the variance of momentum of electrons that is readily measurable by angle-resolved photoemission spectroscopy (ARPES).

We then turn to the crystalline systems to demonstrate that despite a lattice is not a smooth manifold, from the overlap of two such local states at neighboring lattice sites, one can still define a real space quantum metric by means of central difference. Using lattice models of metals and topological insulators, we investigate the metric in the presence of impurities, and calculate the spatial profile of differential geometrical quantities induced near the impurity sites, such as Ricci scalar and volume form similar to those derived from momentum space quantum metric\cite{Matsuura10,Ma13,Ma14,Yang15,Piechon16,Panahiyan20_fidelity,vonGersdorff21_metric_curvature,Mera22,
Kolodrubetz13,Kolodrubetz17,Smith22,Chen25_quantum_geometry_TI_TSC}. This method quantifies the deformation of real space manifold by the impurity, and particularly for 2D systems, it introduces an Euler characteristic that describes the topology of the 2D manifold in the presence of disorder. Our formalism thus demonstrates the possibility of engineering the real space quantum geometry by impurities, which should be broadly applicable to a wide variety of disordered lattice models.



\section{Local state formalism}

\subsection{Single-particle local state \label{sec:single_particle_local_state}}

Consider a solid described by a real space Hamiltonian $H$. Through diagonalizing the Hamiltonian $H|E_{\ell}\rangle=E_{\ell}|E_{\ell}\rangle$, we obtain the eigenstates $|E_{\ell}\rangle$ and eigenenergies $E_{\ell}$, whose wave function $\tilde{a}_{\ell\sigma}({\bf x})$ and local density $n_{\sigma}({\bf x})$ for species $\sigma$ (spin, orbital, sublattice, etc) at position ${\bf x}$ are calculated from the position ket state $|{\bf x},\sigma\rangle$ by
\begin{eqnarray}
&&\langle {\bf x},\sigma|E_{\ell}\rangle=\tilde{a}_{\ell\sigma}({\bf x}),\;\;\;
n_{\sigma}({\bf x})=\sum_{n}|\tilde{a}_{n\sigma}({\bf x})|^{2},
\label{wavefn_atilde_nisigma}
\end{eqnarray}
where we reserve the subscript $n$ for the filled states at zero temperature $E_{n}<0$, and likewisely for the summation $\sum_{n}\equiv\sum_{E_{n}<0}$. Our formalism concerns the normalized wave function $a_{n\sigma}({\bf x})$ calculated from the wave function divided by the square root of the local density
\begin{eqnarray}
a_{n\sigma}({\bf x})\equiv\frac{\tilde{a}_{n\sigma}({\bf x})}{\sqrt{n_{\sigma}({\bf x})}}=\frac{\langle{\bf x},\sigma|E_{n}\rangle}{\sqrt{n_{\sigma}({\bf x})}},\;\;\;\sum_{n}|a_{n\sigma}({\bf x})|^{2}=1.\;\;\;
\end{eqnarray}
The purpose of our work is to elaborate that the notion of quantum metric can be introduced to real space via constructing a local quantum state. This is done by utilizing the projector to the filled lattice eigenstates 
\begin{eqnarray}
P=\sum_{n}|E_{n}\rangle\langle E_{n}|,
\label{projector_P}
\end{eqnarray}
from which we construct a local state by applying the projector to the position and species $|{\bf x},\sigma\rangle$
\begin{eqnarray}
|\psi_{\sigma}({\bf x})\rangle\equiv\frac{1}{\sqrt{n_{\sigma}({\bf x})}}P|{\bf x},\sigma\rangle=\sum_{n}a_{n\sigma}^{\ast}({\bf x})|E_{n}\rangle.\;\;\;
\label{psival_sigmacomplex}
\end{eqnarray}
such that $\langle E_{n}|\psi_{\sigma}({\bf x})\rangle=a_{n\sigma}^{\ast}({\bf x})$. As a result, we see that $|\psi_{\sigma}({\bf x})\rangle$ is a normalized state defined in the Hilbert space spanned by the single-particle eigenstates $|E_{\ell}\rangle$ with coefficients $a_{n\sigma}^{\ast}({\bf x})$, and it does not evolve with time since none of the $\left\{P,n_{\sigma}({\bf x}),|{\bf x},\sigma\rangle\right\}$ does.
This quantum state should be relatively localized around ${\bf x}$, as can be seen by calculating the wave function of this state at some other position ${\bf x'}$ 
\begin{eqnarray}
\langle{\bf x'},\sigma '|\psi_{\sigma}({\bf x})\rangle=\frac{\langle{\bf x'},\sigma '|P|{\bf x},\sigma\rangle}{\sqrt{n_{\sigma}({\bf x})}}, 
\label{local_state_wave_fn}
\end{eqnarray}
which has been argued to decay with $|{\bf x-x'}|$ by viewing the numerator $\langle{\bf x'},\sigma '|P|{\bf x},\sigma\rangle$ as a kind of correlator\cite{Kohn96,Resta06,Thonhauser06,Resta11,Bianco11}. Alternatively, we can also see the overlap of two such states that are a certain distance apart, which gives a similar answer
\begin{eqnarray}
\langle\psi_{\sigma '}({\bf x'})|\psi_{\sigma}({\bf x})\rangle=\frac{\langle{\bf x'},\sigma '|P|{\bf x},\sigma\rangle}{\sqrt{n_{\sigma '}({\bf x'})\,n_{\sigma}({\bf x})}}, 
\label{overlap_psi_xxp}
\end{eqnarray}
and hence should also decay with $|{\bf x-x'}|$. We should see the localization of this state in the examples in the following sections.

Although $|\psi_{\sigma}({\bf x})\rangle$ is not an eigenstate of the Hamiltonian, it serves as a tool to partition expectation values to each position. To see this, consider any one-body operator $\hat{\cal O}$ of a noninteracting system, whose ground state expectation value satisfies
\begin{eqnarray}
&&\sum_{n}\langle E_{n}|\hat{\cal O}|E_{n}\rangle=\sum_{\bf x\sigma}n_{\sigma}({\bf x})\langle\psi_{\sigma}({\bf x})|\hat{\cal O}|\psi_{\sigma}({\bf x})\rangle.
\end{eqnarray}
meaning that the ground state expectation value $\sum_{n}\langle E_{n}|\hat{\cal O}|E_{n}\rangle$ can be partitioned to the species $\sigma$ at position ${\bf x}$ by $\langle\psi_{\sigma}({\bf x})|\hat{\cal O}|\psi_{\sigma}({\bf x})\rangle$ times the local density $n_{\sigma}({\bf x})$. This observation gives a concrete physical interpretation to the local state $|\psi_{\sigma}({\bf x})\rangle$, indicating that it is not just an abstract concept. For instance, if the operator is the Hamiltonian $\hat{\cal O}=H$, then the ground state energy of the Fermi sea $\sum_{n}E_{n}$ has been partitioned by $|\psi_{\sigma}({\bf x})\rangle$ into a local energy resided at $({\bf x},\sigma)$. Finally, it should be reminded again that although this state is normalize correctly $\langle\psi_{\sigma}({\bf x})|\psi_{\sigma}({\bf x})\rangle=1$, it is not an eigenstate of the lattice Hamiltonian $H$ and does not form a complete basis $\sum_{\bf x\sigma}|\psi_{\sigma}({\bf x})\rangle\langle\psi_{\sigma}({\bf x})|\neq I$. The completeness relation is either $\sum_{\bf x\sigma}|{\bf x},\sigma\rangle\langle{\bf x},\sigma|=I$ or $\sum_{\ell}|E_{\ell}\rangle\langle E_{\ell}|=I$.






\subsection{Real space quantum metric}

\begin{figure}[ht]
\begin{center}
\includegraphics[clip=true,width=0.7\columnwidth]{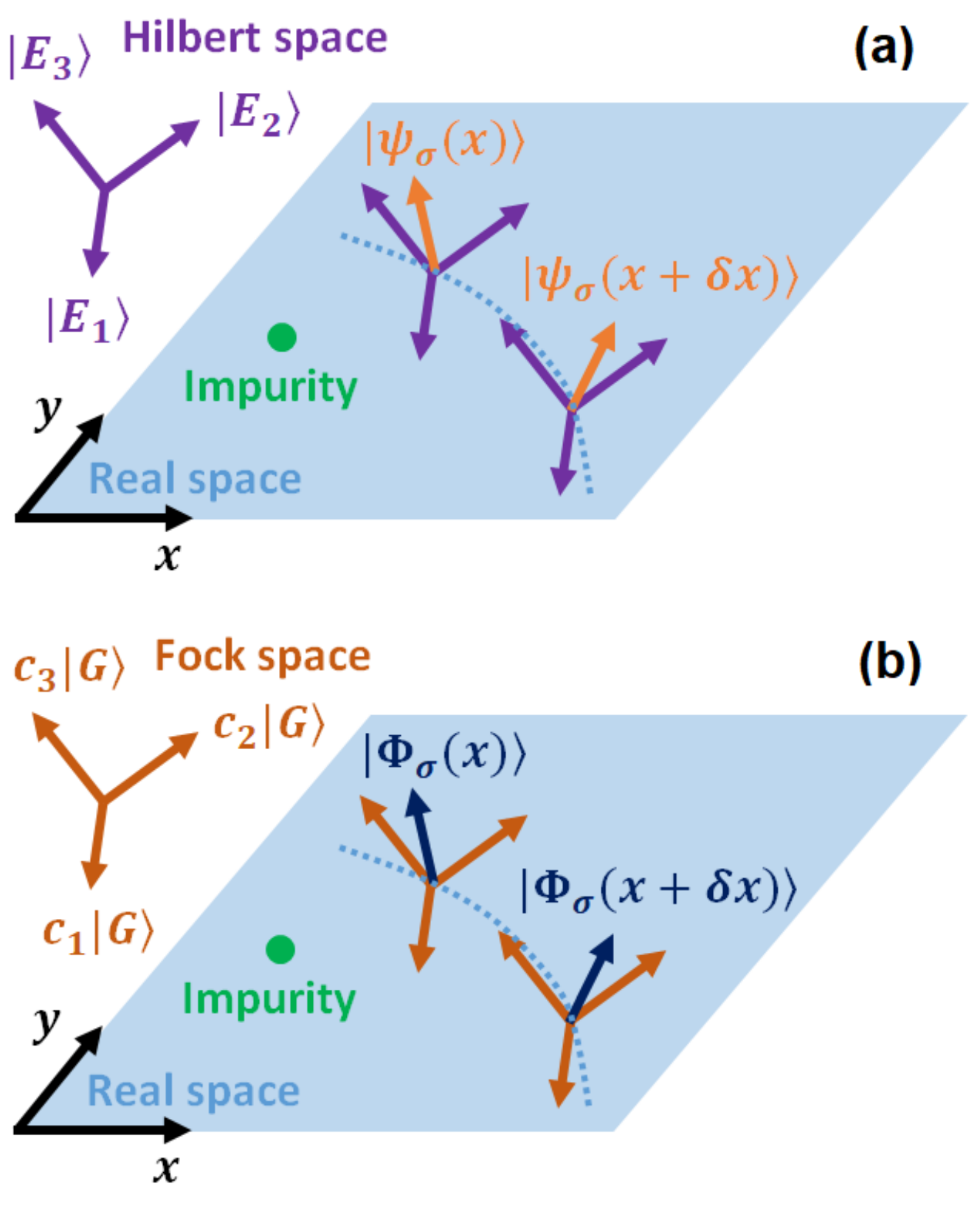}
\caption{Schematics for two ways to construct quantum geometry in real space that yield the same quantum metric: (a) The local state $|\psi_{\sigma}({\bf x})\rangle$ obtained from acting the filled state projector $P$ on a specific position $|{\bf x},\gamma\rangle$, which lives in the Hilbert space spanned by single-particle lattice eigenstates $|E_{\ell}\rangle$. (b) The local state $|\Phi_{\sigma}({\bf x})\rangle$ constructed from acting the local annihilation operator $c_{\bf x\sigma}$ on the many-body ground state $|G\rangle$, which lives in the Fock space spanned by the basis $c_{\ell}|G\rangle$. As one moves from ${\bf x}$ to ${\bf x+\delta x}$ in the real space of a disordered system, $|\psi_{\sigma}({\bf x})\rangle$ as a unit vector will rotate in the Hilbert space and $|\Phi_{\sigma}({\bf x})\rangle$ as a unit vector will rotate in the Fock space, and the overlap of unit vectors $|\langle\psi_{\sigma}({\bf x})|\psi_{\sigma}({\bf x+\delta x})\rangle|$ and $|\langle\Phi_{\sigma}({\bf x})|\Phi_{\sigma}({\bf x+\delta x})\rangle|$ define the quantum metric $g_{\mu\nu}$.  } 
\label{fig:quantum_metric_schematics}
\end{center}
\end{figure}

We proceed to define a quantum metric $g_{\mu\nu}$ from
the overlap of two local states at slightly different locations\cite{Provost80}
\begin{eqnarray}
&&|\langle\psi_{\sigma}({\bf x})|\psi_{\sigma}({\bf x+\delta x})\rangle|=\left|\sum_{n}a_{n\sigma}({\bf x})a_{n\sigma}^{\ast}({\bf x+\delta x})\right|
\nonumber \\
&&=1-\frac{1}{2}g_{\mu\nu}({\bf x},\sigma)\delta x^{\mu}\delta x^{\nu},
\label{gmunu_single_particle_definition}
\end{eqnarray}
that measures the distance between ${\bf x}$ and ${\bf x+\delta x}$ specifically for the species $\sigma$. The physical picture for the metric is given in Fig.~\ref{fig:quantum_metric_schematics} (a), where the local state $|\psi_{\sigma}({\bf x})\rangle$ is schematically a unit vector in the Hilbert space spanned by the energy eigenstate $\left\{|E_{\ell}\rangle\right\}$. The unit vector changes with position ${\bf x}$, and the quantum metric measures the product of the unit vectors $|\psi_{\sigma}({\bf x})\rangle$ and $|\psi_{\sigma}({\bf x+\delta x})\rangle$ at slightly different positions. Intuitively, if the wave function of the local state $\langle{\bf x'},\sigma '|\psi_{\sigma}({\bf x})\rangle$ and that at a slightly different position $\langle{\bf x'},\sigma '|\psi_{\sigma}({\bf x+\delta x})\rangle$ have a small overlap, then the metric is large, and vice versa. Hence the metric is also a measure of the extendedness of the wave function, as indicated schematically in Fig.~\ref{fig:gmunu_overlap_schematics}. In Appendix \ref{apx:vielbein_like_formalism}, we further introduce a matrix product-like form for the metric, and discuss the issue of quantum geometric tensor and Berry curvature.





\begin{figure}[ht]
\begin{center}
\includegraphics[clip=true,width=0.99\columnwidth]{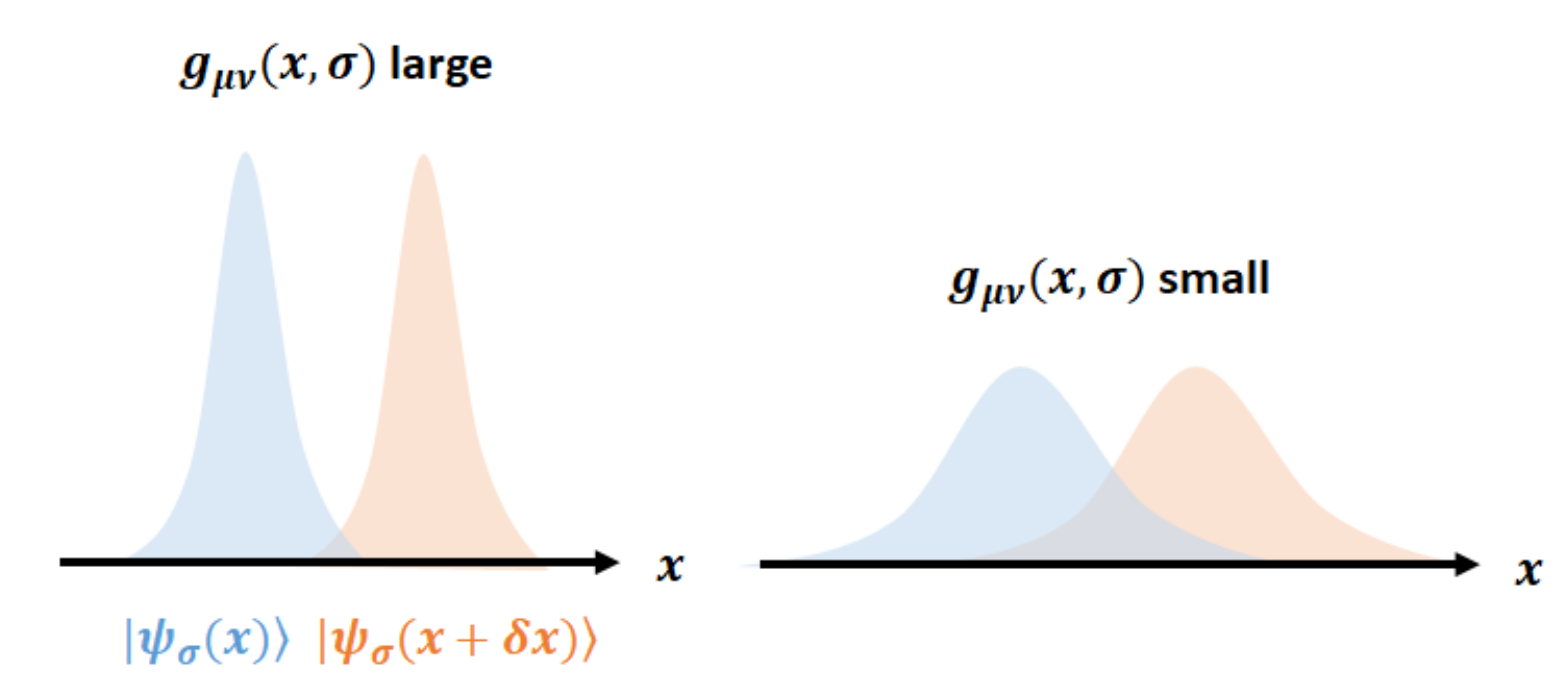}
\caption{Schematics of the real space quantum metric $g_{\mu\nu}({\bf x},\sigma)$. If the wave function of the local state $\langle{\bf x'},\sigma '|\psi_{\sigma}({\bf x})\rangle$ is very localized, then the overlap $|\langle\psi_{\sigma}({\bf x})|\psi_{\sigma}({\bf x+\delta x})\rangle|$ is small and hence the quantum metric is large (left panel). In contrast, if $\langle{\bf x'},\sigma '|\psi_{\sigma}({\bf x})\rangle$ is very extended, then the overlap is large, rendering a small quantum metric (right panel). } 
\label{fig:gmunu_overlap_schematics}
\end{center}
\end{figure}

Having the real space quantum metric $g_{\mu\nu}$ defined, one can further calculate the differential geometrical quantities that describe the real space as a Euclidean manifold, such as the Christoffel symbol $\Gamma_{\mu\nu}^{\lambda}$, Riemann tensor $R^{\rho}_{\;\xi\mu\nu}$, Ricci Tensor $R_{\mu\nu}$, Ricci scalar $R$, and Einstein tensor $G_{\mu\nu}$ defined by\cite{Carroll19}
\begin{eqnarray}
&&\Gamma_{\mu\nu}^{\lambda}=\frac{1}{2}g^{\lambda\xi}(\partial_{\mu}g_{\nu\xi}
+\partial_{\nu}g_{\xi\mu}-\partial_{\xi}g_{\mu\nu}),
\nonumber \\
&&R^{\rho}_{\;\xi\mu\nu}=\partial_{\mu}\Gamma_{\nu\xi}^{\rho}-\partial_{\nu}\Gamma_{\mu\xi}^{\rho}
+\Gamma_{\mu\lambda}^{\rho}\Gamma_{\nu\xi}^{\lambda}-\Gamma_{\nu\lambda}^{\rho}\Gamma_{\mu\xi}^{\lambda},
\nonumber \\
&&R_{\mu\nu}=R^{\lambda}_{\;\mu\lambda\nu},\;\;\;
R=g^{\mu\nu}R_{\nu\mu},
\nonumber \\
&&G_{\mu\nu}=R_{\mu\nu}-\frac{1}{2}Rg_{\mu\nu},
\label{Christoffel_Riemann_Ricci}
\end{eqnarray} 
that may be derived in a unified manner from a generating function\cite{Hetenyi23}, where $g^{\mu\nu}$ is the inverse matrix of $g_{\mu\nu}$ that satisfies $g_{\mu\nu}g^{\nu\rho}=\delta_{\mu}^{\rho}$, and the covariant derivative is defined with respect to the real space coordinates $\partial_{\mu}=\partial/\partial x^{\mu}$. Particularly for 2D systems with a periodic boundary condition, one can introduce the Euler characteristic for the compact $T^{2}$ manifold
\begin{eqnarray}
\chi=\int\frac{d^{2}{\bf x}}{4\pi}\sqrt{g}\,R,
\label{Euler_characteristic}
\end{eqnarray}
a concept that has been heavily exploited in the momentum space quantum geometry of topological materials\cite{Matsuura10,Ma13,Ma14,Yang15,Chen25_quantum_geometry_TI_TSC}. For the real space quantum geometry proposed in the present work, one expects the Euler characteristic to remain an integer even in the presence of disorder, as we aim to demonstrate.

\subsection{Density-averaged quantum metric}

Given that the formalism in the previous section defines the quantum metric $g_{\mu\nu}({\bf x},\sigma)$ for each degree of freedom $\sigma$ separately, it is intriguing to ask if there is a way to combine all the degrees of freedom together to define a single characteristic metric. We suggest the following construction based on the connection to experiments that will be demonstrated in Sec.~\ref{sec:general_homogeneous_solid}. We observe that the density for each $\sigma$ relative to the total density is the probability $P_{\sigma}({\bf x})$ that an electron at position ${\bf x}$ has a specific degree of freedom $\sigma$
\begin{eqnarray}
P_{\sigma}({\bf x})=\frac{n_{\sigma}({\bf x})}{\sum_{\sigma}n_{\sigma}({\bf x})},
\end{eqnarray}
satisfying $\sum_{\sigma}P_{\sigma}({\bf x})=1$. Thus one can define a density-averaged metric by weighting each $\sigma$ by the probability and summing them together
\begin{eqnarray}
\overline{g}_{\mu\nu}({\bf x})=\sum_{\sigma}P_{\sigma}({\bf x})g_{\mu\nu}({\bf x},\sigma).
\label{density_averaged_T_g_Omega}
\end{eqnarray}
This is equivalent to defining a single characteristic metric from a density-averaged overlap of the local states
\begin{eqnarray}
\sum_{\sigma}P_{\sigma}({\bf x})|\langle\psi_{\sigma}({\bf x})|\psi_{\sigma}({\bf x+\delta x})\rangle|=1-\frac{1}{2}\overline{g}_{\mu\nu}({\bf x})\delta x^{\mu}\delta x^{\nu}.
\nonumber \\
\end{eqnarray}
We shall see the significance of $\overline{g}_{\mu\nu}$ in Sec.~\ref{sec:general_homogeneous_solid}.

\subsection{Real space quantum metric defined from a many-body local state}

In the section, we elaborate yet another interpretation of quantum metric from a localized many-body state $|\Phi_{\sigma}({\bf x})\rangle$. As we shall see below, the local state $|\Phi_{\sigma}({\bf x})\rangle$ can be viewed as a unit vector in the many-body Fock space of multiple Fermions, which is different from the single-particle local state $|\psi_{\sigma}({\bf x})\rangle$ introduced in Sec.~\ref{sec:single_particle_local_state} that is a unit vector in the single-particle Hilbert space. Nevertheless, the two interpretations give the same quantum metric in practice. We start by considering the ground state $|G\rangle$ of the system given by
\begin{eqnarray}
|G\rangle=\prod_{n=1}^{N_{-}}c_{n}^{\dag}|0\rangle=|1_{1},1_{2}...1_{N_{-}}\rangle,
\label{many_body_ground_state}
\end{eqnarray}
where $c_{n}^{\dag}$ is the electron creation operator for eigenstate $|E_{n}\rangle$, $|0\rangle$ is the vacuum state that contains no electrons, the label $1_{n}$ indicates that the $n$-th eigenstate is filled with one electron, and totally there are $N_{-}$ filled eigenstates. We consider a local state constructed by applying the electron annihilation operator $c_{\bf x\sigma}$ for species $\sigma$ at position ${\bf x}$ to the ground state
\begin{eqnarray}
&&c_{\bf x\sigma}|G\rangle
=\sum_{\ell}\tilde{a}_{\ell\sigma}({\bf x})c_{\ell}|G\rangle
=\sum_{n}\tilde{a}_{n\sigma}({\bf x})c_{n}|G\rangle
\nonumber \\
&&=\sum_{n}\tilde{a}_{n\sigma}({\bf x})(-1)^{\sum_{n'<n}1}|1_{1}...0_{n}...1_{N_{-}}\rangle,
\end{eqnarray}
where we have changed the basis $c_{\bf x\sigma}=\sum_{\ell}\tilde{a}_{\ell\sigma}({\bf x})c_{\ell}$, and noted that only the electron in the filled eigenstate $|E_{n}\rangle$ can be annihilated, and the sign $(-1)^{\sum_{n'<n}1}$ is due to $\left\{c_{n},c_{n'}\right\}=0$. Because this state is not normalized correctly $\langle G|c_{\bf x\sigma}^{\dag}c_{\bf x\sigma}|G\rangle=\sum_{n}|a_{n\sigma}({\bf x})|^{2}=n_{\sigma}({\bf x})$ due to $\langle G|c_{n'}^{\dag}c_{n}|G\rangle=\delta_{nn'}$, we further introduce a correctly normalized local state by dividing by the density $n_{\sigma}({\bf x})$
\begin{eqnarray}
|\Phi_{\sigma}({\bf x})\rangle\equiv\frac{1}{\sqrt{n_{\sigma}({\bf x})}}\,c_{\bf x\sigma}|G\rangle=\sum_{n}a_{n\sigma}({\bf x})\,c_{n}|G\rangle,
\end{eqnarray}
which satisfies $\langle\Phi_{\sigma}({\bf x})|\Phi_{\sigma}({\bf x})\rangle=1$. The state $|\Phi_{\sigma}({\bf x})\rangle$ is therefore a unit vector defined in the Fock space spanned by the basis $c_{\ell}|G\rangle$. We can further see the localization of this state by considering the overlap of two such states at different locations
\begin{eqnarray}
&&\langle\Phi_{\sigma '}({\bf x'})|\Phi_{\sigma}({\bf x})\rangle
=\sum_{n}a_{n\sigma '}^{\ast}({\bf x'})a_{n\sigma}({\bf x})
\nonumber \\
&&=\langle\psi_{\sigma '}({\bf x'})|\psi_{\sigma}({\bf x})\rangle^{\ast},
\end{eqnarray}
which is equal to the complex conjugate of the overlap of the single-particle local state given in Eq.~(\ref{overlap_psi_xxp}), and hence also decays with $|{\bf x-x'}|$, signifying the localization of $|\Phi_{\sigma}({\bf x})\rangle$.

The overlap of the many-body local state $|\Phi_{\sigma}({\bf x})\rangle$ defines a quantum metric identical to that introduced from the single-particle local state $|\psi_{\sigma}({\bf x})\rangle$ in Eq.~(\ref{gmunu_single_particle_definition}), since
\begin{eqnarray}
&&|\langle\Phi_{\sigma}({\bf x})|\Phi_{\sigma}({\bf x+\delta x})\rangle|=\left|\sum_{n}a_{n\sigma}^{\ast}({\bf x})a_{n\sigma}({\bf x+\delta x})\right|
\nonumber \\
&&=1-\frac{1}{2}g_{\mu\nu}({\bf x},\sigma)\delta x^{\mu}\delta x^{\nu}.
\label{quantum_metric_expression_many_body}
\end{eqnarray}
It should be reminded that the physical meaning of the state $|\psi_{\sigma}({\bf x})\rangle$ is not the same as $|\Phi_{\sigma}({\bf x})\rangle$, since the latter is a unit vector defined in the many-body Fock space spanned by $\left\{c_{n}|G\rangle\right\}$, as shown schematically in Fig.~\ref{fig:quantum_metric_schematics} (b). Nevertheless, the resulting quantum metric defined from the overlap of the single-particle local state $|\langle \psi_{\sigma}({\bf x})|\psi_{\sigma}({\bf x+\delta x})\rangle|$ and from that of the many-particle local state $|\langle\Phi_{\sigma}({\bf x})|\Phi_{\sigma}({\bf x+\delta x})\rangle|$ are the same quantity.

Note that the quantity $\langle G|c_{\bf x\sigma}^{\dag}c_{\bf x'\sigma}|G\rangle$ evaluated at two arbitrary positions ${\bf x}$ and ${\bf x}'$ is precisely the single-particle correlation function of electrons (see the discussion about pair distribution function in Ref.~\onlinecite{Mahan00}). For instance, in 3D homogeneous Fermi gas it is given by the particle density times a spherical Bessel function. Thus our many-body formalism of quantum metric via Eq.~(\ref{quantum_metric_expression_many_body}) is essentially the expansion of the density-normalized single-particle correlation function $\langle G|c_{\bf x\sigma}^{\dag}c_{\bf x+\delta x\sigma}|G\rangle/\sqrt{n_{\sigma}({\bf x})n_{\sigma}({\bf x+\delta x})}$ at small distance $\delta {\bf x}$, thereby giving this correlation function a quantum geometrical interpretation. 





\section{Continuous models} 

We now address the continuous systems where the position ${\bf x}$ is a continuous variable. In these systems, the displacement in Eq.~(\ref{gmunu_single_particle_definition}) can be infinitesimal $\delta{\bf x}\rightarrow 0$, and an expansion over $\delta{\bf x}$ gives the quantum metric
\begin{eqnarray}
&&g_{\mu\nu}({\bf x},\sigma)=\frac{1}{2}\sum_{n}\partial_{\mu}a_{n\sigma}\partial_{\nu}a_{n\sigma}^{\ast}
+\frac{1}{2}\sum_{n}\partial_{\nu}a_{n\sigma}\partial_{\mu}a_{n\sigma}^{\ast}
\nonumber \\
&&-\left[\sum_{n}\left(\partial_{\mu}a_{n\sigma}\right)a_{n\sigma}^{\ast}\right]
\left[\sum_{n'}a_{n'\sigma}\partial_{\nu}a_{n'\sigma}^{\ast}\right]
\label{quantum_metric_expression}
\end{eqnarray}
We will use the particles in a box and homogeneous electron gas to explicitly calculate the metric.

\subsection{Particles in a box}

Our first example are particles in a box, which, despite their simplicity, may be relevant to realistic mesoscopic systems like quantum dots. Consider spinless fermionic particles in a $D$-dimensional rectangular box of volume $V=L_{1}L_{2}...L_{D}$, assuming $n_{\rm max}$ of the discrete eigenstates are filled at zero temperature due to Pauli exclusion principle. The coordinates in each direction is confined within $0<x^{\mu}<L_{\mu}$, and we omit the subscript $\sigma$ since there is only one degree of freedom. The wave functions are
\begin{eqnarray}
\langle{\bf x}|E_{n}\rangle=\frac{2^{D/2}}{\sqrt{V}}\prod_{\rho=1}^{D}\sin\frac{n_{\rho}\pi x^{\rho}}{L_{\rho}}, 
\end{eqnarray}
and hence the normalized local state in Eq.~(\ref{psival_sigmacomplex}) is
\begin{eqnarray}
|\psi({\bf x})\rangle=\frac{\sum_{n}|E_{n}\rangle\prod_{\nu=1}^{D}\sin\frac{n_{\nu}\pi x^{\nu}}{L_{\nu}}}{\sqrt{\sum_{n'}\prod_{\rho=1}^{D}\sin^{2}\left(\frac{n_{\rho}^{\prime}\pi x^{\rho}}{L_{\rho}}\right)}},
\end{eqnarray}
where the quantized wave vector in each direction for each eigenstate is characterized by the set of integers
\begin{eqnarray}
&&{\bf n}_{1}=(n_{1x},n_{1y}...n_{1D})=(1,1...1),
\nonumber \\
&&{\bf n}_{2}=(n_{2x},n_{2y}...n_{2D})=(1,1...2),
\nonumber \\
&&\vdots
\nonumber \\
&&{\bf n}_{\rm max}=(n_{\rm max,x},n_{\rm max,y}...n_{\rm max,D}).
\end{eqnarray}
Using the vielbein formalism in Eqs.~(\ref{enu_uvec}) and (\ref{gmunu_vielbein}), we introduce the vector ${\hat{\bf u}}$ of normalized wave function 
\begin{eqnarray}
{\hat{\bf u}}=\frac{2^{D/2}}{\sqrt{Vn_{\rm max}}}\left(\begin{array}{c}
\prod_{\rho=1}^{D}\sin\frac{n_{1\rho}\pi x^{\rho}}{L_{\rho}} \\
\prod_{\rho=1}^{D}\sin\frac{n_{2\rho}\pi x^{\rho}}{L_{\rho}} \\
\vdots \\
\prod_{\rho=1}^{D}\sin\frac{n_{\rm max\rho}\pi x^{\rho}}{L_{\rho}} \\
\end{array}\right),
\end{eqnarray}
Because the wave functions are real, the quantum metric is the same as the quantum geometric tensor, given by
\begin{eqnarray}
&&g_{\mu\nu}=\langle\partial_{\mu}\psi|\partial_{\nu}\psi\rangle
=\partial_{\mu}{\hat{\bf u}}\cdot\partial_{\nu}{\hat{\bf u}}=T_{\mu\nu}.
\end{eqnarray}
Note that $g_{\mu\nu}=0$ if only the lowest eigenstate ${\bf n}_{1}=(1,1...1)$ is filled, such as in the case of a single particle or bosons at zero temperature.

\begin{figure}[ht]
\begin{center}
\includegraphics[clip=true,width=0.7\columnwidth]{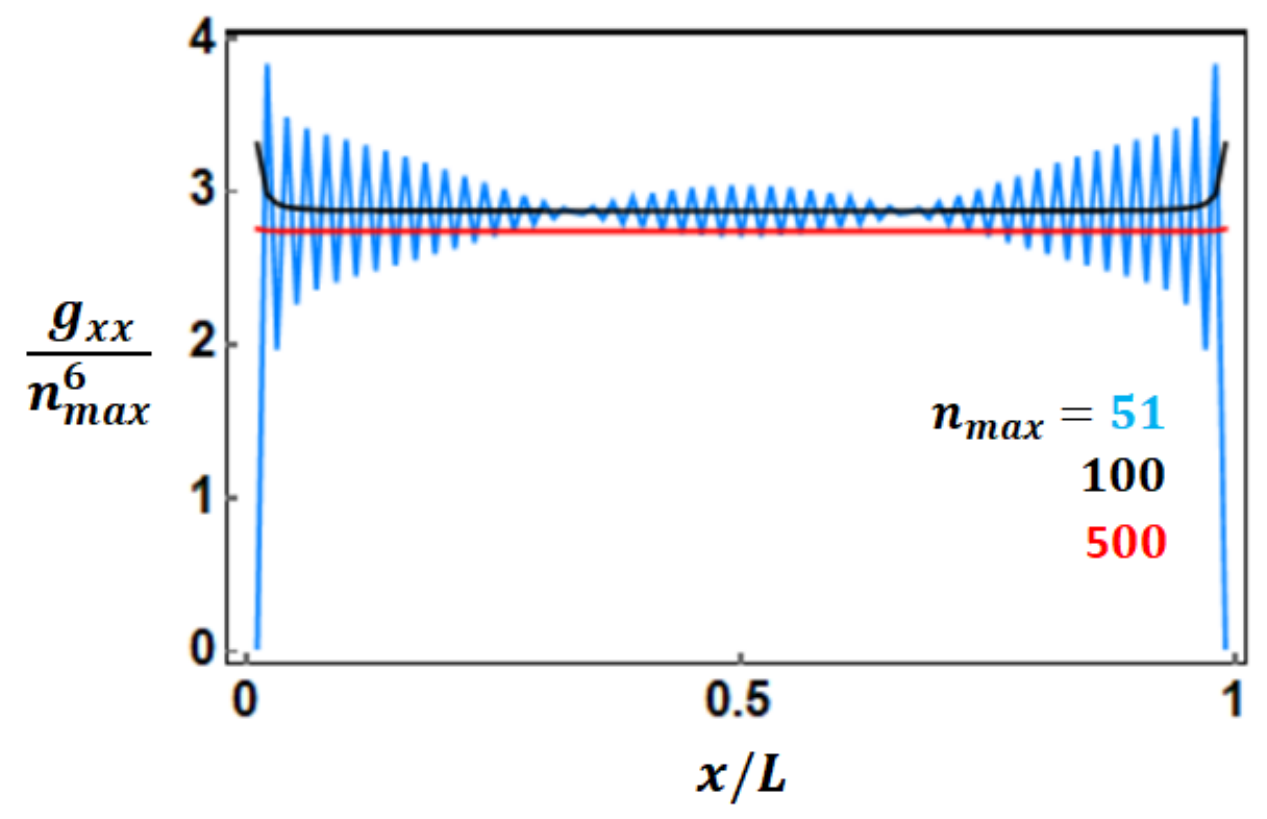}
\caption{The spatial profile of quantum metric $g_{xx}$ for fermionic particles in a 1D box, which depends on the number of particles $n_{\rm max}$. We find that $g_{xx}/n_{\rm max}^{6}$ in a large area saturates to about $2.87$ at large values of $n_{\rm max}$, indicating $g_{xx}$ scales like $n_{\rm max}^{6}$. } 
\label{fig:particles_in_box_1D}
\end{center}
\end{figure}

The spatial profile of $g_{\mu\nu}$ depends on the number of filled states $n_{\rm max}$. Figure \ref{fig:particles_in_box_1D} shows the spatial profile of $g_{xx}$ in the 1D box $D=1$, which is simplified to
\begin{eqnarray}
g_{xx}&=&\frac{\left[\sum_{n}\cos^{2}\left(\frac{n\pi x}{L}\right)\left(\frac{n\pi}{L}\right)^{2}\right]}{\sum_{n'}\sin^{2}\left(\frac{n'\pi x}{L}\right)}
\nonumber \\
&&-\frac{\left[\sum_{n}\sin\left(\frac{n\pi x}{L}\right)\cos\left(\frac{n\pi x}{L}\right)\left(\frac{n\pi}{L}\right)\right]^{2}}{\left[\sum_{n'}\sin^{2}\left(\frac{n'\pi x}{L}\right)\right]^{2}}.
\end{eqnarray}
We find that at large values of $n_{\rm max}$, $g_{xx}$ scales like $n_{\rm max}$ to the sixth power. The value of $g_{xx}/n_{\rm max}^{6}$ in the most part of $0<x<L$ saturates to about 2.87 as $n_{\rm max}\rightarrow\infty$. The case of higher dimensional boxes $D>1$ and the related geometrical quantities in Eq.~(\ref{Christoffel_Riemann_Ricci}), as well as quantum dots with irregular shapes, should be left for future investigations.

\subsection{General formalism for homogeneous solids \label{sec:general_homogeneous_solid}}

We now address the real space quantum metric in a homogeneous and continuous solid with periodic boundary condition. In this case, the momentum ${\bf k}$ remains a good quantum number and the density is a constant $n_{\sigma}({\bf x})=n_{\sigma}$, so the local state constructed by the projector $P$ is equivalently 
\begin{eqnarray}
&&|\psi_{\sigma}({\bf x})\rangle=\frac{1}{\sqrt{n_{\sigma}}}\int\frac{d^{D}{\bf k}}{(2\pi\hbar/a)^{D}}\sum_{n}|\phi_{n{\bf k}}\rangle\langle\phi_{n{\bf k}}|{\bf x},\sigma\rangle
\nonumber \\
&&=\frac{1}{\sqrt{n_{\sigma}}}\int\frac{d^{D}{\bf k}}{(2\pi\hbar/a)^{D}}\sum_{n}|\phi_{n{\bf k}}\rangle\,e^{-i{\bf k\cdot x}/\hbar}\tilde{a}_{n\sigma}^{\ast}({\bf k}),
\end{eqnarray}
where the full state $|\phi_{n{\bf k}}\rangle$ is related to the periodic part of the Bloch state $|n_{\bf k}\rangle$ by
\begin{eqnarray}
\langle{\bf x},\sigma|\phi_{n{\bf k}}\rangle=e^{i{\bf k\cdot x}/\hbar}\langle\sigma|n_{\bf k}\rangle=e^{i{\bf k\cdot x}/\hbar}\tilde{a}_{n\sigma}({\bf k}),
\end{eqnarray}
and we have imposed a fictitious lattice constant $a$ to regularize the integration despite the position ${\bf x}$ is continuous. The Bloch state $|n_{\bf k}\rangle$ and its eigenenergy $\varepsilon_{\bf k}$ are obtained from a momentum space Hamiltonian $H_{\bf k}$ via $H_{\bf k}|n_{\bf k}\rangle=\varepsilon_{\bf k}|n_{\bf k}\rangle$. In practice, the Bloch state is a column
\begin{eqnarray}
|n_{\bf k}\rangle=\left(\begin{array}{c}
\tilde{a}_{n1}({\bf k}) \\
\tilde{a}_{n2}({\bf k}) \\
\vdots
\end{array}\right)
\end{eqnarray}
whose entries are the wave function $\tilde{a}_{n\sigma}({\bf k})$ for the degree of freedom $\sigma$. The derivative on the local state then gives
\begin{eqnarray}
|\partial_{\mu}\psi_{\sigma}({\bf x})\rangle&=&\frac{1}{\sqrt{n_{\sigma}}}\int\frac{d^{D}{\bf k}}{(2\pi\hbar/a)^{D}}
\nonumber \\
&&\times\sum_{n}|\phi_{n{\bf k}}\rangle\left(-\frac{i}{\hbar}k_{\mu}\right)e^{-i{\bf k\cdot x}/\hbar}\tilde{a}_{n\sigma}^{\ast}({\bf k}),\;\;\;
\end{eqnarray}
and therefore the quantum metric and the quantum geometric tensor are the same, given by 
\begin{eqnarray}
&&g_{\mu\nu}({\bf x},\sigma)=\langle\partial_{\mu}\psi_{\sigma}({\bf x})|\partial_{\nu}\psi_{\sigma}({\bf x})\rangle
\nonumber \\
&&=\frac{1}{n_{\sigma}}\int\frac{d^{D}{\bf k}}{(2\pi\hbar/a)^{D}}
\frac{k_{\mu}k_{\nu}}{\hbar^{2}}\sum_{n}|\tilde{a}_{n\sigma}({\bf k})|^{2},
\label{gmunu_homogeneous}
\end{eqnarray}
since $\langle\partial_{\mu}\psi_{\sigma}({\bf x})|\psi_{\sigma}({\bf x})\rangle=0$ vanishes due to oddness of the integrand in the momentum integration. 




\subsection{ARPES measurement of homogeneous quantum metric}

The homogeneous quantum metric for continuous systems formulated in the previous section can be measured by ARPES, and has a remarkable physical significance as the momentum variance of electrons. To see this, we start from the retarded Green's function of the $n$-th band Bloch state $|n_{\bf k}\rangle$
\begin{eqnarray}
G_{n}({\bf k},\omega)=\frac{|n_{\bf k}\rangle\langle n_{\bf k}|}{\omega-\varepsilon_{n{\bf k}}/\hbar+i\eta},
\end{eqnarray}
which is a matrix over the degrees of freedom $\sigma$ due to the $|n_{\bf k}\rangle\langle n_{\bf k}|$ part, and $\eta$ is a small artificial broadening. The single-particle spectral function of the $n$-th band is given by
\begin{eqnarray}
&&A_{n}({\bf k},\omega)=-\frac{1}{\pi}{\rm Im\;Tr}\,G_{n}({\bf k},\omega)
\nonumber \\
&&=\sum_{\sigma}|\tilde{a}_{n\sigma}({\bf k})|^{2}\delta(\omega-\varepsilon_{n{\bf k}}/\hbar)=\delta(\omega-\varepsilon_{n{\bf k}}/\hbar),
\end{eqnarray}
where the trace is over $\sigma$, yielding a $\delta$-function that peaks at $\omega=\varepsilon_{n{\bf k}}/\hbar$ in noninteracting systems. As a result, the frequency-integration of the spectra function gives the particle number of the $n$-th Bloch state at momentum ${\bf k}$
\begin{eqnarray}
\sum_{\sigma}|\tilde{a}_{n\sigma}({\bf k})|^{2}=\langle n_{\bf k}|n_{\bf k}\rangle=\int_{-\infty}^{0}d\omega\,A_{n}({\bf k},\omega)=1,\;\;\;
\label{int_Ankw_1}
\end{eqnarray}
and consequently the homogeneous particle density is 
\begin{eqnarray}
&&\sum_{\sigma}n_{\sigma}=\int\frac{d^{D}{\bf k}}{(2\pi\hbar/a)^{D}}\sum_{n}\sum_{\sigma}|\tilde{a}_{n\sigma}({\bf k})|^{2}
\nonumber \\
&&=\int\frac{d^{D}{\bf k}}{(2\pi\hbar/a)^{D}}\int_{-\infty}^{0}d\omega\sum_{n}A_{n}({\bf k},\omega)
\nonumber \\
&&=\int\frac{d^{D}{\bf k}}{(2\pi\hbar/a)^{D}}\int_{-\infty}^{0}d\omega\,A({\bf k},\omega),
\end{eqnarray}
where in the last line we have introduced the ARPES spectral function that is measured experimentally
\begin{eqnarray}
A({\bf k},\omega)=\sum_{n}A_{n}({\bf k},\omega),
\label{Akw_Ankw_equivalence}
\end{eqnarray}
that sums over all the filled bands. Comparing with Eq.~(\ref{density_averaged_T_g_Omega}), we see that the density-averaged quantum metric $\overline{g}_{\mu\nu}$ can be expressed by the spectral function
\begin{eqnarray}
&&\overline{g}_{\mu\nu}
=\frac{\int\frac{d^{D}{\bf k}}{(2\pi\hbar/a)^{D}}\frac{k_{\mu}k_{\nu}}{\hbar^{2}}\int_{-\infty}^{0}d\omega\,A({\bf k},\omega)}{\int\frac{d^{D}{\bf k}}{(2\pi\hbar/a)^{D}}\int_{-\infty}^{0}d\omega\,A({\bf k},\omega)}
=\left\langle\frac{k_{\mu}k_{\nu}}{\hbar^{2}}\right\rangle.\;\;\;\;\;
\label{density_averaged_metric}
\end{eqnarray}
This expression provides a concrete protocol to measure the density-averaged quantum metric $\overline{g}_{\mu\nu}$ from ARPES spectral function: It is simply the momentum-frequency integration of ARPES spectral function $A({\bf k},\omega)$ weighted by the momentum variance $k_{\mu}k_{\nu}/\hbar^{2}$ and normalized by particle density. In addition, because $A({\bf k},\omega)$ normalized by density is precisely the probability of finding an electron with momentum ${\bf k}$ and energy $\hbar\omega$, from the usual definition in the probability theory, $\overline{g}_{\mu\nu}$ also has the physical meaning as the variance (or covariance) of momentum $\langle k_{\mu}k_{\nu}/\hbar^{2}\rangle$ of the electrons. Provided $A({\bf k},\omega)$ is measured in the whole BZ and down to the bottom of the bands for either metals or insulators, which is easily accessible by ARPES experiments, $\overline{g}_{\mu\nu}$ can be readily extracted by Eq.~(\ref{density_averaged_metric}). Moreover, the trace ${\rm Tr}\,\overline{g}_{\mu\nu}=\langle k^{2}/\hbar^{2}\rangle$ is invariant under the rotation of coordinates, thereby serving as a characteristic feature of the Fermi sea. Finally, we remark that the covariance $\langle k_{\mu}k_{\nu}/\hbar^{2}\rangle$ bears a striking resemblance with the spatial components of the energy-momentum tensor of noninteracting particles in general relativity, as it can be interpreted as the flux of the $\mu$-direction momentum flowing along $\nu$-direction\cite{Carroll19}, manifesting an appealing connection to general relativity.

We now discuss several issues one may encounter in a real ARPES experiment. Firstly, the above analysis does not take into account many features of the real ARPES spectral function $A_{n}^{\rm exp}({\bf k},\omega)$ in experiments. Firstly, $A_{n}^{\rm exp}({\bf k},\omega)$ is in general not a $\delta$-function in frequency due to the scattering caused by, e.g., phonons and disorder. Moreover, $A_{n}^{\rm exp}({\bf k},\omega)$ in reality is weighted by the matrix element that describes the excitation of Bloch electrons in a solid to the free electrons in vacuum measured by the detector, and hence the relation in Eq.~(\ref{int_Ankw_1}) is in general not valid $\int_{-\infty}^{0}d\omega\,A_{n}^{\rm exp}({\bf k},\omega)\neq 1$. To cope with these realistic factors, we suggest that one may simply approximate the experimental ARPES spectral function by a $\delta$-function $A_{n}^{\rm exp}({\bf k},\omega)\rightarrow\delta(\omega-\varepsilon_{n{\bf k}})$ that peaks at the maximum $\partial A_{n}^{\rm exp}({\bf k},\omega)/\partial\omega|_{\omega=\varepsilon_{n{\bf k}}}=0$, and then the above analysis follows. Note that this is how the band structure $\varepsilon_{n{\bf k}}$ is usually determined experimentally anyway, and we simply approximate the spectral function by a $\delta$-function accordingly. The momentum variance $\langle k_{\mu}k_{\nu}/\hbar^{2}\rangle$ can then be calculated by Eq.~(\ref{density_averaged_metric}), which should be a good estimation for the measured material provided that many-body interactions are not too strong.



\subsection{Homogeneous electron gases}

The simplest example of homogeneous systems is a $D$-dimensional spinless electron gas that has only one degree of freedom, where one can ignore $\sigma$ and set $\tilde{a}_{n\sigma}({\bf k})=1$. The wave function of its local state is calculated from
\begin{eqnarray}
&&\int\frac{d^{D}{\bf k}}{(2\pi\hbar/a)^{D}}\sum_{n}\langle{\bf x}|\phi_{n{\bf k}}\rangle\langle\phi_{n{\bf k}}|{\bf 0}\rangle
=\int_{FS}\frac{d^{D}{\bf k}}{(2\pi\hbar/a)^{D}}e^{i{\bf k\cdot x}/\hbar}
\nonumber \\
&&=\left\{\begin{array}{l}
\frac{\sin k_{F}x/\hbar}{\pi x/a}\;\;\;(D=1) \\
\frac{1-\cos k_{F}r/\hbar}{2\pi r^{2}/a^{2}}\;\;\;(D=2) \\
\frac{1}{(2\pi\hbar/a)^{3}}\left[-\frac{4\pi\hbar^{2}k_{F}}{r^{2}}\cos\frac{k_{F}r}{\hbar}
+\frac{4\pi\hbar^{3}}{r^{3}}\sin\frac{k_{F}r}{\hbar}\right]\;\;\;(D=3) 
\end{array}
\right.
\nonumber \\
\label{plane_wave_overlap}
\end{eqnarray}
where $\int_{FS}$ indicates the integration over all the plane wave states inside a $D$-dimensional sphere of radius $k_{F}$ in momentum space, and we have  assigned a lattice constant $a$ to obtain the correct unit. Comparing Eq.~(\ref{plane_wave_overlap}) with Eq.~(\ref{local_state_wave_fn}), the oscillating and decaying behavior of the wave function of the local state $\langle{\bf x}|\psi({\bf 0})\rangle$ with respect to $x$ is evident, indicating that $|\psi({\bf 0})\rangle$ is genuinely localized at the origin ${\bf 0}$. Furthermore, the homogeneous particle number is 
\begin{eqnarray}
&&n_{0}=\int_{FS}\frac{d^{D}{\bf k}}{(2\pi\hbar/a)^{D}}
=\left\{\begin{array}{l}
\frac{1}{\pi}\left(\frac{k_{F}}{\hbar/a}\right)\;\;\;(D=1) \\
\frac{1}{4\pi}\left(\frac{k_{F}}{\hbar/a}\right)^{2}\;\;\;(D=2) \\
\frac{1}{6\pi^{2}}\left(\frac{k_{F}}{\hbar/a}\right)^{3}\;\;\;(D=3) \\
\end{array}
\right.\;\;\;
\end{eqnarray} 
The quantum metric in the Cartesian coordinates can be calculated from Eq.~(\ref{gmunu_homogeneous}), rendering
\begin{eqnarray}
g_{\mu\mu}=(g^{\mu\mu})^{-1}=\frac{1}{D+2}\left(\frac{k_{F}}{\hbar}\right)^{2}
\end{eqnarray}
for $1\leq D\leq 3$, while the off-diagonal elements vanish $g_{\mu\neq\nu}=0$. Thus $g_{\mu\nu}$ is simply given by the square of the Fermi momentum as expected, since it is the characteristic length scale in this problem. Note that because the Bloch state in momentum space $|u({\bf k})\rangle$ is simply a trivial unity, there is no momentum space quantum metric in this case since $|\langle u({\bf k})|u({\bf k+\delta k})\rangle|=1$. Nevertheless, the real space quantum metric still exists, indicating that it is a more ubiquitous concept compared to the momentum space quantum metric. 



\section{Crystalline systems \label{sec:disordered_lattice}}

We proceed to use tight-binding models to investigate the quantum metric in crystalline systems where the position ${\bf x}$ takes discrete values due to the crystalline structure, and address how the profile of the metric is modified under the influence of disorder. For this purpose, we consider a $D$-dimensional orthorhombic, tetragonal, or cubic lattice with unit vectors ${\bf a}_{\mu}=a_{\mu}{\hat{\boldsymbol\mu}}$ that are orthogonal to each other, where $\mu=1\sim D$ denotes the spatial directions. Lattices belonging to other point groups should be addressed elsewhere. The unit cells are located at Bravais lattice vectors ${\bf x}=\sum_{\mu=1}^{D}n_{\mu}{\bf a}_{\mu}$, where $n_{\mu}$ are integers. The degrees of freedom in a unit cell is labeled by $\sigma$, which may signify spin, orbit, sublattice, etc. The lattice Hamiltonian takes the following generic form
\begin{eqnarray}
H=\sum_{\bf xx'\sigma\sigma '}t_{\bf xx'\sigma\sigma '}c_{\bf x,\sigma}^{\dag}c_{\bf x',\sigma '},
\end{eqnarray}
where $c_{\bf x,\sigma}^{\dag}$ is the creation operator of electron of species $\sigma$ at site ${\bf x}$.

As elaborated in Appendix \ref{apx:implementing_metric_on_lattice}, even in disordered crystalline systems, the metric and differential geometrical properties can still be implemented on discrete lattice points provided the derivatives in Eqs.~(\ref{quantum_metric_expression}) and (\ref{Christoffel_Riemann_Ricci}) are calculated numerically via central difference, as demonstrated below by a 2D metal and the Chern insulator. Moreover, in disordered systems, $\overline{g}_{\mu\nu}({\bf x})$ generally varies with position ${\bf x}$ in the atomic scale, as we shall see in the following sections. It remains unclear to us at present whether this atomic scale variation can be experimentally detected by certain local probes.

\subsection{Wannier state representation in crystalline insulators and semiconductors}

In homogeneous and fully gapped crystalline insulators and semiconductors, the real space quantum metric can be expressed in terms of the overlap of neighboring Wannier functions. To see this, we denote $\langle{\bf r},\sigma|\ell_{\bf k}\rangle=\ell_{\bf k\sigma}({\bf r})=e^{-i{\bf k\cdot r}}\phi_{\ell{\bf k}\sigma}({\bf r})$ as the periodic part of the Bloch state satisfying $\ell_{\bf k\sigma}({\bf r})=\ell_{\bf k\sigma}({\bf r+R})$, where ${\bf r}$ and ${\bf R}$ are Bravais lattice vectors and ${\bf k}$ is the momentum, and introduce the Wannier state $|{\bf R}\ell\rangle$ by
\begin{eqnarray}
|\ell_{\bf k}\rangle=\sum_{{\bf R}}e^{-i {\bf k}\cdot({\hat{\bf r}}-{\bf R})/\hbar}|{\bf R}\ell\rangle,\;\;
|{\bf R} \ell\rangle=\sum_{\bf k}e^{i {\bf k}\cdot({\hat{\bf r}}-{\bf R})/\hbar}|\ell_{\bf k}\rangle,
\nonumber \\
\label{Wannier_basis}
\end{eqnarray}
where all the degrees of freedom inside a unit cell are assigned with the same Bravais lattice vector\cite{Vanderbilt18}. The corresponding Wannier function $\langle {\bf x},\sigma|{\bf R} \ell\rangle=W_{\ell\sigma}({\bf x}-{\bf R})$ at position ${\bf x}$ and species $\sigma$ is highly localized around the home cell ${\bf R}$. Denoting $\sum_{n}$ as the summation over the filled valence band states, the local state in Eq.~(\ref{psival_sigmacomplex}) in this case is\cite{Marzari97}
\begin{eqnarray}
&&|\psi_{\sigma}({\bf x})\rangle=\frac{1}{\sqrt{n_{\sigma}({\bf x})}}\sum_{n{\bf k}}|\phi_{n{\bf k}}\rangle\langle\phi_{n{\bf k}}|{\bf x},\sigma\rangle
\nonumber \\
&&=\frac{1}{\sqrt{n_{\sigma}({\bf x})}}\sum_{n{\bf R}}|{\bf R}n\rangle\langle{\bf R}n|{\bf x},\sigma\rangle
\nonumber \\
&&=\sum_{n{\bf R}}|{\bf R}n\rangle\frac{W_{n\sigma}^{\ast}({\bf x-R})}{\sqrt{n_{\sigma}({\bf x})}}.
\end{eqnarray}
From which it follows the overlap of the local states
\begin{eqnarray}
&&|\langle\psi_{\sigma}({\bf x})|\psi_{\sigma}({\bf x+\delta x})\rangle|^{2}
\nonumber \\
&&=\frac{|\sum_{n{\bf R}}W_{n\sigma}({\bf x-R})W_{n\sigma}^{\ast}({\bf x+\delta x-R})|^{2}}{n_{\sigma}({\bf x})n_{\sigma}({\bf x+\delta x})}.
\label{overlap_Wannier_representation}
\end{eqnarray}
This expression gives the quantum metric $g_{\mu\nu}$ in Eq.~(\ref{gmunu_single_particle_definition}) an interpretation in terms of the valence band Wannier functions: The overlap evaluates the product of the values at $({\bf x},\sigma)$ of the two neighboring Wannier functions centering at ${\bf R}$ and at ${\bf R-\delta x}$, and then summing this product over all home cells ${\bf R}$. The quantum metric then measures how much this summation of products deviates from particle density $n_{\sigma}({\bf x})n_{\sigma}({\bf x+\delta x})$.

\begin{figure*}[ht]
\begin{center}
\includegraphics[clip=true,width=1.9\columnwidth]{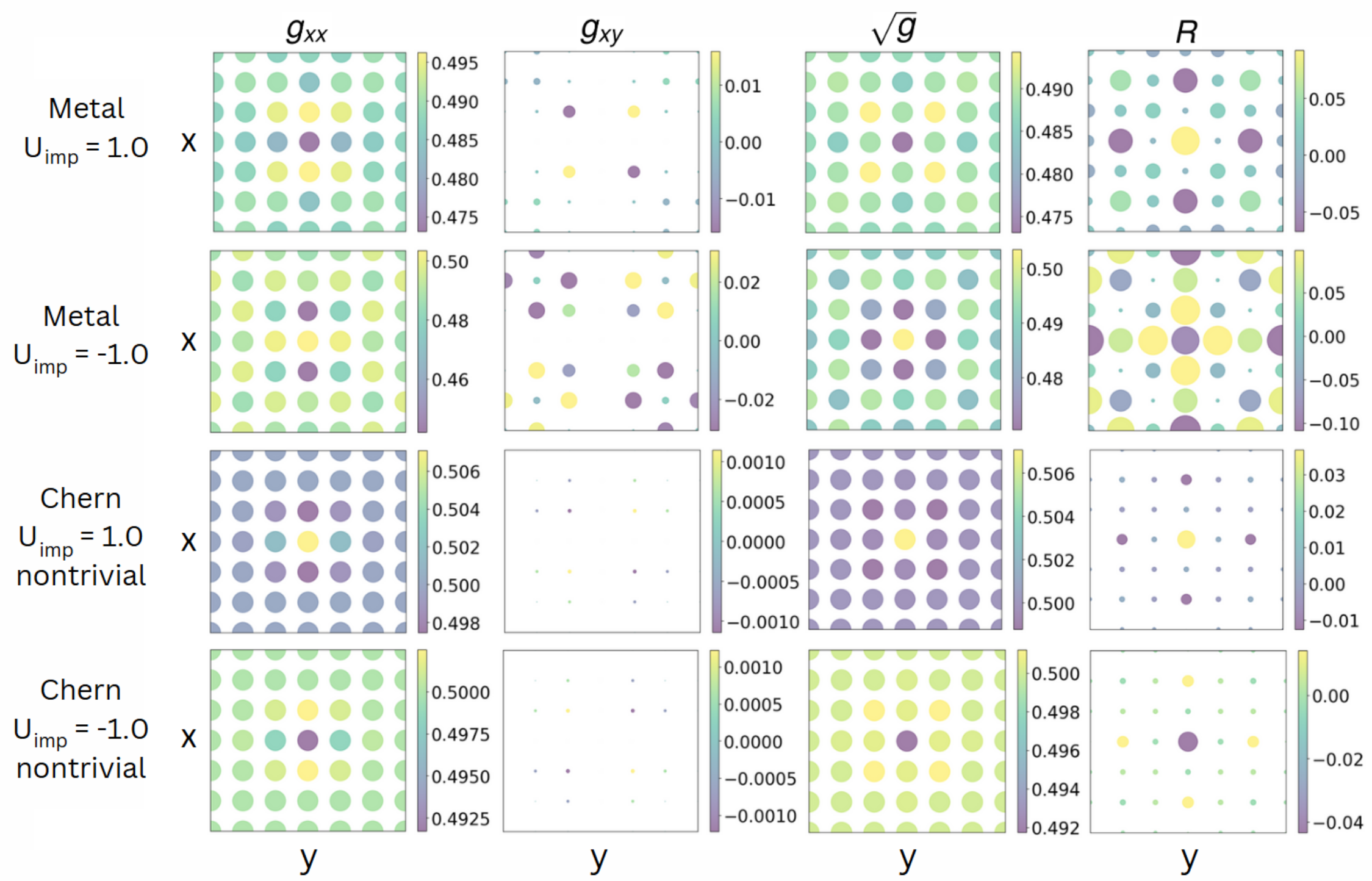}
\caption{The spatial pattern of the (density-averaged) diagonal $g_{xx}$ and off-diagonal $g_{xy}$ components of real space quantum metric, the volume form $\sqrt{g}$ (all in units of inverse of lattice constant square $1/a^{2}$), and the Ricci scalar $R$ (unitless) induced by an impurity potential $U_{imp}$ located at the center of the 2D lattice, investigated for four different systems from top to bottom: 2D metal at chemical potential $\mu=-0.1$ with positive $U_{imp}=1$ and negative $U_{imp}=-1$, and Chern insulator in the topologically nontrivial phase $M=-2$ with positive $U_{imp}=1$ and negative $U_{imp}=-1$. Note that in both systems, the Ricci scalar on the impurity site $R({\bf x}_{imp})$ changes from positive (yellow) at $U_{imp}=1$ to negative (blue) at $U_{imp}=-1$.  } 
\label{fig:2Dmetal_Chern_figure}
\end{center}
\end{figure*}

\subsection{Lattice model of a 2D metal}

The first lattice model under consideration are 2D spinless fermions on a square lattice with nearest-neighbor hopping and a single impurity at the origin
\begin{eqnarray}
H=-t\sum_{\bf x,a}\left(c_{\bf x}^{\dag}c_{\bf x+a}+c_{\bf x+a}^{\dag}c_{\bf x}\right)
-\mu\sum_{\bf x}c_{\bf x}^{\dag}c_{\bf x}+U_{imp}\,c_{\bf 0}^{\dag}c_{\bf 0}.
\nonumber \\
\end{eqnarray}
where ${\bf x}=(x,y)$ labels the 2D lattice sites, and ${\bf a}=\left\{{\bf a}^{x},{\bf a}^{y}\right\}$ the lattice vectors in the two directions. We set $t=1$, and investigate different chemical potential $\mu$ and impurity potential $U_{imp}$.


The numerical results for the diagonal $g_{xx}$ and off-diagonal $g_{xy}$ components of the metric, the volume form $\sqrt{g}$, and the Ricci scalar $R$ at chemical potential $\mu=-0.1$ are shown in Fig.~\ref{fig:2Dmetal_Chern_figure}. At impurity potential $U_{imp}=1$ and $U_{imp}=-1$, the absolute scale of the diagonal component of the metric $g_{xx}$ is of the order of unity in units of the inverse of lattice constant square $1/a^{2}$, and it is not much influenced by the presence of the impurity. Even at strong impurity potential like $U_{imp}=10$, the variation of $g_{xx}$ around the impurity site is only few percent. However, the pattern of the variation strongly depends on $U_{imp}$, and the pattern is generally not four-fold symmetric but mirror-symmetric around the impurity site. In addition, the impurity induces a small off-diagonal component $g_{xy}$, which are of opposite signs along the two diagonal crystalline directions. Nevertheless, the resulting volume form $\sqrt{g}$ is four-fold symmetric around the impurity site, and we find that the variation of $g_{\mu\nu}$ and $\sqrt{g}$ generally becomes more long ranged as the impurity strength $U_{imp}$ increases, and oscillates in a way that mimics the Friedel oscillation. Finally, the Ricci scalar $R$ shows a four-fold symmetric pattern that spatially oscillates between positive and negative values, indicating a modulation of positive and negative curvature depending on the distance away from the impurity.

\begin{figure}[ht]
\begin{center}
\includegraphics[clip=true,width=0.99\columnwidth]{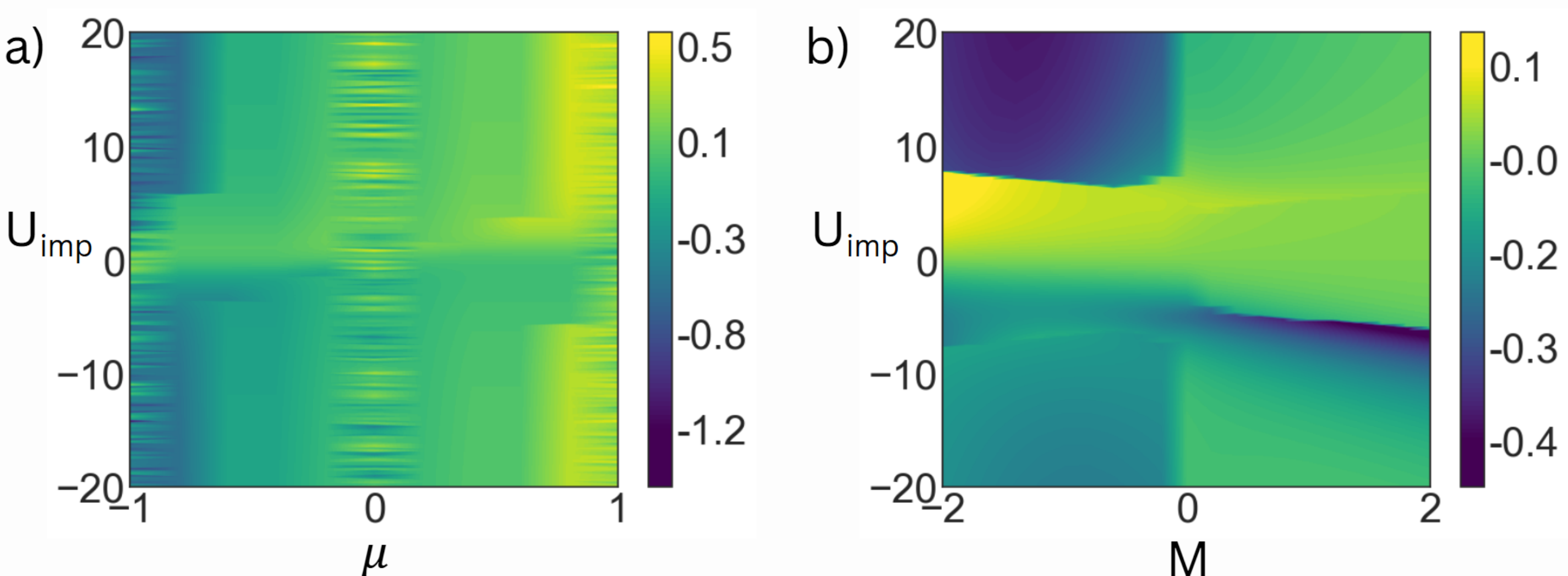}
\caption{The Ricci scalar on the impurity site $R({\bf x}_{imp})$ for (a) the normal metal as a function of chemical potential $\mu$ and impurity potential $U_{imp}$, and (b) the Chern insulator as a function the mass term $M$ and impurity potential $U_{imp}$.  } 
\label{fig:phase_diagram_Rimp}
\end{center}
\end{figure}

Interestingly, the Ricci scalar is found to be strongly depending on the parameters, suggesting the feasibility of engineering the curvature by disorder. To be more specific, the sign of Ricci scalar on the impurity site $R({\bf x}_{imp})$ depends strongly on both the chemical potential $\mu$ and impurity potential $U_{imp}$, as can be seen by comparing the $U_{imp}=1$ and $U_{imp}=-1$ results in Fig.~\ref{fig:2Dmetal_Chern_figure}. The full dependence of $R({\bf x}_{imp})$ on $\mu$ and $U_{imp}$ is further quantified in Fig.~\ref{fig:phase_diagram_Rimp}. Intuitively, a positive or negative Ricci scalar on the impurity site is geometrically analogous to a hump or a depression on an otherwise flat surface, and our results indicate that it can be controlled by $\mu$ and $U_{imp}$. The spatial integration of $\sqrt{g}\,R$ always gives a zero Euler characteristic $\chi=0$ (up to numerical precision) at any parameter, indicating that a single impurity does not change the topology of the manifold. This is in a way analogous to the fact that a local hump or depression on the surface does not change the number of handles of a large 3D object.




\subsection{Lattice model of Chern insulator}

As an example of insulators, we examine the lattice model of Chern insulators with a single impurity given by\cite{Chen20_absence_edge_current,Molignini23_Chern_marker} 
\begin{eqnarray}
&&H=\sum_{\bf x}t\left\{-ic_{{\bf x}s}^{\dag}c_{{\bf x+a}^{x}p}
+ic_{{\bf x+a}^{x}s}^{\dag}c_{{\bf x}p}+h.c.\right\}
\nonumber \\
&&+\sum_{\bf x}t\left\{-c_{{\bf x}s}^{\dag}c_{{\bf x+a}^{y}p}+c_{{\bf x+a}^{y}s}^{\dag}c_{{\bf x}p}+h.c.\right\}
\nonumber \\
&&+\sum_{{\bf x}}\sum_{\delta=\left\{x,y\right\}}t'\left\{-c_{{\bf x}s}^{\dag}c_{{\bf x+a}^{\delta} s}+c_{{\bf x}p}^{\dag}c_{{\bf x+a}^{\delta} p}+h.c.\right\}
\nonumber \\
&&+\sum_{\bf x}\left(M+4t'\right)\left\{c_{{\bf x}s}^{\dag}c_{{\bf x}s}
-c_{{\bf x}p}^{\dag}c_{{\bf x}p}\right\},
\nonumber \\
&&+U_{imp}\left\{c_{{\bf 0}s}^{\dag}c_{{\bf 0}s}
+c_{{\bf 0}p}^{\dag}c_{{\bf 0}p}\right\},
\label{Hamiltonian_2DclassA}
\end{eqnarray} 
where $\sigma=\left\{s,p\right\}$ are the orbitals, and the impurity potential is assumed to be the same for the two orbitals on the impurity site. The effect of disorder has been an intriguing issue for this model, especially whether the topological invariant remains unchanged in the presence of disorder, which can be investigated by the Chern marker\cite{Bianco11,Prodan10,Costa19,Ulcakar20,dOrnellas22,Oliveira24_impurity_marker} that is also constructed from the projector in Eq.~(\ref{projector_P}).

We consider the parameters $t=1.0$, $t'=1.0$, and $M=-2$ for the topologically nontrivial phase, and calculate the density-averaged matric $\overline{g}_{\mu\nu}$ and the resulting differential geometrical quantities as given in Fig.~\ref{fig:2Dmetal_Chern_figure}. Once again the diagonal element $g_{xx}$ is of the order of unity, and is not much influenced by the presence of impurity, yet the impurity does induce a small off-diagonal element $g_{xy}$. These quantities, as well as the volume form $\sqrt{g}$, decay rapidly to the homogeneous value as moving away from the impurity site without much oscillation, which seem to inherit from the fact that insulators do not have a characteristic length scale other than the lattice constant itself. The resulting Ricci scalar $R$ is relatively short ranged compared to the metals for the same reason. Once again we find that the spatial integration of $\sqrt{g}\,R$ is zero up to numerical precision, suggesting that the Euler characteristic always remains zero $\chi=0$ in the presence of the single impurity.

On the other hand, the sign of the Ricci scalar $R$ is found to strongly depend on the mass term $M$ that controls the momentum space topological order (the Chern number) and the impurity potential $U_{imp}$, as can be seen by comparing the $U_{imp}=1$ and $U_{imp}=-1$ results in Fig.~\ref{fig:2Dmetal_Chern_figure}. The full dependence of the impurity site Ricci scalar $R({\bf x}_{imp})$ on $M$ and $U_{imp}$ is shown in Fig.~\ref{fig:phase_diagram_Rimp}, where we see that the sign of $R({\bf x}_{imp})$ strongly depends on whether the system is in the topologically nontrivial $M<0$ or trivial $M>0$ phases, as well as the magnitude and sign of $U_{imp}$, implying the possibility of engineering the real space quantum geometry by disorder and system parameters. Intuitively, this is analogous to creating a local hump or depression on the real space 2D manifold by means of adjusting $M$ and $U_{imp}$, in a way that the number of handles $\chi=0$ remains unchanged.


\section{Conclusions}

In summary, we propose a local state formalism that allows to introduce the notion of quantum geometry to the real space of solids. We observe that by applying the projector to filled states on a specific lattice site, or by applying the electron annihilation operator on the ground state, one can construct a quantum state whose wave function is localized around the designated lattice site. The overlap of two such states at neighboring lattice sites defines a real space quantum metric, from which one can introduce various differential geometrical quantities like Riemann tensor and Ricci scalar to characterize the real space Euclidean manifold. In continuous and homogeneous systems, the density-averaged quantum metric is given by the momentum variance of electrons, indicating the possibility of experimentally detect the metric by ARPES. In crystalline systems, the presence of disorder distorts the spatial profile of the metric and deforms the manifold, offering the possibility of engineering the real space quantum geometry by disorder. We anticipate that our formalism should be ubiquitously applicable to any fermionic systems in arbitrary dimensions, and may be generalized to investigate more exotic systems like superconductors or strongly correlated systems, which would enable the investigation of quantum geometry under the influence of both inhomogeneity and interactions. These interesting issues await to be further explored.

\appendix

\section{Properties of the real space quantum geometric tensor \label{apx:vielbein_like_formalism}}

The normalized local state $|\psi_{\sigma}({\bf x})\rangle$ in Eq.~(\ref{psival_sigmacomplex}) allows to introduce the quantum geometric tensor $T_{\mu\nu}$, whose real part is the quantum metric $g_{\mu\nu}$, and we denote its imaginary part by $\Omega_{\mu\nu}$, which have the generic expressions in continuous systems when the displacement $\delta{\bf x}\rightarrow 0$ is infinitesimal
\begin{eqnarray}
T_{\mu\nu}({\bf x},\sigma)&=&\langle\partial_{\mu}\psi_{\sigma}|\partial_{\nu}\psi_{\sigma}\rangle
-\langle\partial_{\mu}\psi_{\sigma}|\psi_{\sigma}\rangle\langle\psi_{\sigma}|\partial_{\nu}\psi_{\sigma}\rangle,
\nonumber \\
&=&g_{\mu\nu}({\bf x},\sigma)-\frac{i}{2}\Omega_{\mu\nu}({\bf x},\sigma),
\nonumber \\
g_{\mu\nu}({\bf x},\sigma)&=&\frac{1}{2}\langle\partial_{\mu}\psi_{\sigma}|\partial_{\nu}\psi_{\sigma}\rangle
+\frac{1}{2}\langle\partial_{\nu}\psi_{\sigma}|\partial_{\mu}\psi_{\sigma}\rangle
\nonumber \\
&&-\langle\partial_{\mu}\psi_{\sigma}|\psi_{\sigma}\rangle\langle\psi_{\sigma}|\partial_{\nu}\psi_{\sigma}\rangle,
\nonumber \\
\Omega_{\mu\nu}({\bf x},\sigma)&=&i\partial_{\mu}\psi_{\sigma}|\partial_{\nu}\psi_{\sigma}\rangle
-i\langle\partial_{\nu}\psi_{\sigma}|\partial_{\mu}\psi_{\sigma}\rangle.
\label{Tmunu_gmunu_Omegamunu}
\end{eqnarray}
We can always put these quantities into matrix product forms by introducing a two-column matrix
\begin{eqnarray}
&&e_{\nu}({\bf x},\sigma)=\left(|\psi_{\sigma}\rangle,|\partial_{\nu}\psi_{\sigma}\rangle\right)
=\left(\begin{array}{cc}
\psi_{\sigma,1} & \partial_{\nu}\psi_{\sigma,1} \\
\psi_{\sigma,2} & \partial_{\nu}\psi_{\sigma,2} \\
\vdots & \vdots
\end{array}\right),
\nonumber \\
&&\det e_{\mu}^{\dag}e_{\nu}=\det\left(\begin{array}{cc}
\langle\psi_{\sigma}|\psi_{\sigma}\rangle & \langle\psi_{\sigma}|\partial_{\nu}\psi_{\sigma}\rangle \\
\langle\partial_{\mu}\psi_{\sigma}|\psi_{\sigma}\rangle & \langle\partial_{\mu}\psi_{\sigma}|\partial_{\nu}\psi_{\sigma}\rangle \\
\end{array}\right),
\label{enu_detenu}
\end{eqnarray}
where $\psi_{\sigma i}$ are the entries to the vector $|\psi_{\sigma}\rangle$, the quantities in Eq.~(\ref{Tmunu_gmunu_Omegamunu}) can be written as matrix products
\begin{eqnarray}
&&T_{\mu\nu}(\sigma)=\det e_{\mu}^{\dag}e_{\nu},\;\;\;
\nonumber \\
&&g_{\mu\nu}(\sigma)={\rm Re}\left[\det e_{\mu}^{\dag}e_{\nu}\right],
\nonumber \\
&&\Omega_{\mu\nu}(\sigma)=-2\,{\rm Im}\left[\det e_{\mu}^{\dag}e_{\nu}\right].
\end{eqnarray} 
Note that this formalism is generic and applied to other properly normalized quantum state $|\psi({\boldsymbol\lambda})\rangle$ depending on whatever continuous parameter ${\boldsymbol\lambda}$, in which case $\partial_{\mu}=\partial/\partial\lambda^{\mu}$.

Alternative to the formula in Eq.~(\ref{enu_detenu}), for the real space quantum metric introduced in the present work, one may also define a two-column matrix directly from the normalized wave function
\begin{eqnarray}
\varepsilon_{\nu}^{\dag}({\bf x},\sigma)&=&\left(\begin{array}{cc}
a_{1\sigma}^{\ast}({\bf x}) & \partial_{\nu}a_{1\sigma}^{\ast}({\bf x}) \\
a_{2\sigma}^{\ast}({\bf x}) & \partial_{\nu}a_{2\sigma}^{\ast}({\bf x}) \\
\vdots & \vdots \\
a_{n_{\rm max}\sigma}^{\ast}({\bf x}) & \partial_{\nu}a_{n_{\rm max}\sigma}^{\ast}({\bf x}) \\
\end{array}\right)
\equiv \left({\hat{\bf u}}_{\sigma}^{\ast},\partial_{\nu}{\hat{\bf u}}_{\sigma}^{\ast}\right),\;\;\;
\nonumber \\
\label{enu_uvec}
\end{eqnarray}
where the unit vector ${\hat{\bf u}}_{\sigma}^{\ast}$ that satisfies ${\hat{\bf u}}_{\sigma}^{T}\cdot{\hat{\bf u}}_{\sigma}^{\ast}=1$ is simply the array of the normalized wave function $a_{n\sigma}^{\ast}({\bf x})$. In terms of these quantities, one has
\begin{eqnarray}
&&T_{\mu\nu}(\sigma)=\det\left(\begin{array}{cc}
{\hat{\bf u}}_{\sigma}^{T}\cdot{\hat{\bf u}}_{\sigma}^{\ast} & {\hat{\bf u}}_{\sigma}^{T}\cdot\partial_{\nu}{\hat{\bf u}}_{\sigma}^{\ast} \\
\partial_{\mu}{\hat{\bf u}}_{\sigma}^{T}\cdot{\hat{\bf u}}_{\sigma}^{\ast} & \partial_{\mu}{\hat{\bf u}}_{\sigma}^{T}\cdot\partial_{\nu}{\hat{\bf u}}_{\sigma}^{\ast}
\end{array}\right)
\nonumber \\
&&=\det \varepsilon_{\mu}\varepsilon_{\nu}^{\dag},
\nonumber \\
&&g_{\mu\nu}(\sigma)={\rm Re}\left[\det \varepsilon_{\mu}\varepsilon_{\nu}^{\dag}\right],
\nonumber \\
&&\Omega_{\mu\nu}(\sigma)=-2{\rm Im}\left[\det \varepsilon_{\mu}\varepsilon_{\nu}^{\dag}\right],
\label{gmunu_vielbein}
\end{eqnarray}
which also renders a vielbein-like form\cite{Carroll19} in terms of $\varepsilon_{\nu}(\sigma)$. Note that the $\varepsilon_{\nu}(\sigma)$ in Eq.~(\ref{gmunu_vielbein}) is not the same as the $e_{\nu}(\sigma)$ in Eq.~(\ref{enu_detenu}), although they both result in a vielbein-like form for $g_{\mu\nu}$. In practice, whether $\varepsilon_{\nu}(\sigma)$ or $e_{\nu}(\sigma)$ is more useful depends on the problem one encounters in practical calculations. As a final remark, although our real space quantum metric appears somewhat analogous to Einstein's spacetime metric, they have different units. The elements of Einstein's spacetime metric are unitless numbers, while our real space quantum metric has the unit of inverse meter square $\left[g_{\mu\nu}\right]=1/$m$^{2}$ as can be seen from its very definition in Eq.~(\ref{gmunu_single_particle_definition}).

We now remark on the physical interpretation of the quantum geometric tensor $T_{\mu\nu}$ and its imaginary part $\Omega_{\mu\nu}$. Firstly, the overlap in Eq.~(\ref{gmunu_single_particle_definition}) can as well be expressed by $T_{\mu\nu}$
\begin{eqnarray}
&&|\langle\psi_{\sigma}({\bf x})|\psi_{\sigma}({\bf x+\delta x})\rangle|=1-\frac{1}{2}T_{\mu\nu}({\bf x},\sigma)\delta x^{\mu}\delta x^{\nu},
\label{overlap_Tmunu}
\end{eqnarray}
since the imaginary part sums to zero. Hence one may interpret $T_{\mu\nu}$ as a more generalized complex metric. On the other hand, it should be noted that $\Omega_{\mu\nu}$ is {\it not} a Berry curvature as usually defined, since the local state $|\psi_{\sigma}({\bf x})\rangle$ is not an eigenstate of some Hamiltonian. This can be seen by noticing that the very definition of the Berry phase concerns an eigenstate $|n({\cal R})\rangle$ of a Hamiltonian $H({\cal R})$ satisfying $H({\cal R})|n({\cal R})\rangle=E_{n}({\cal R})|n({\cal R})\rangle$ and depending on some external parameter ${\cal R}$. When the parameter ${\cal R}$ is taken along some closed trajectory in the parameter space, the eigenstate accumulates a Berry phase $\gamma_{n}=\oint d{\cal R}\langle n({\cal R})|i\nabla_{\cal R}|n({\cal R})\rangle$ that can be written as the integration of Berry curvature $\nabla_{\cal R}\times\langle n({\cal R})|i\nabla_{\cal R}|n({\cal R})\rangle$ over the area enclosed by the trajectory\cite{Berry84,Xiao10}. Analogously, one may vary the parameter ${\bf x}$ along some closed trajectory in real space and calculate the evolution of our local state $|\psi_{\sigma}({\bf x})\rangle$. However, since $|\psi_{\sigma}({\bf x})\rangle$ is not an eigenstate of some real space Hamiltonian $H({\bf x})$, this process will not generate the Berry phase discussed above, nor does $\Omega_{\mu\nu}$ in Eq.~(\ref{Tmunu_gmunu_Omegamunu}) have the physical meaning as a Berry curvature. Thus $\Omega_{\mu\nu}$ is merely the imaginary part of $T_{\mu\nu}$ defined from the overlap of the local state in Eq.~(\ref{overlap_Tmunu}) that does not have much significance in our theory.

\section{Implementing quantum metric on discrete lattice sites \label{apx:implementing_metric_on_lattice}}

The real space of crystalline systems is not a smooth manifold, since the positions ${\bf x}$ are only defined on discrete lattice sites and the displacement $\delta{\bf x}$ cannot be infinitesimal. Nevertheless, below we elaborate that quantum metric can still be implemented on such a discretized space, using the lattice vectors as the smallest increment possible $\delta{\bf x}={\bf a}$. Firstly, since our local state is correctly normalized $\langle\psi_{\sigma}({\bf x})|\psi_{\sigma}({\bf x})\rangle=1$, the overlap of two such states defined on any two sites ${\bf x}$ and ${\bf x+a}$ must be less than unity
\begin{eqnarray}
|\langle\psi_{\sigma}({\bf x})|\psi_{\sigma}({\bf x+a})\rangle|\leq 1.
\end{eqnarray}
Hence if the separation ${\bf a}$ is sufficiently small, one must be able to expand the overlap by a series of powers of ${\bf a}$. Secondly, if the system is translationally invariant without disorder, then 
\begin{eqnarray}
&&|\langle\psi_{\sigma}({\bf x})|\psi_{\sigma}({\bf x+a})\rangle|=
|\langle\psi_{\sigma}({\bf x-a})|\psi_{\sigma}({\bf x})\rangle|
\nonumber \\
&&=|\langle\psi_{\sigma}({\bf x})|\psi_{\sigma}({\bf x-a})\rangle|,
\end{eqnarray}
so the series must contain only even powers of ${\bf a}$. The leading order is the second order, whose coefficient is then naturally identified as the quantum metric.

To formulate the quantum metric rigorously, we assume a $D$-dimensional orthorhombic, tetragonal, or cubic lattice with lattice vector ${\bf a}^{\mu}=a^{\mu}{\hat{\boldsymbol\mu}}$ in the $\mu$-direction, and employ discrete Taylor expansion with central difference up to second order
\begin{widetext}
\begin{eqnarray}
&&|\psi_{\sigma}({\bf x+a})\rangle=|\psi_{\sigma}({\bf x})\rangle+\frac{1}{2}\sum_{\mu}\left[|\psi_{\sigma}({\bf x+a}^{\mu})\rangle-|\psi_{\sigma}({\bf x-a}^{\mu})\rangle\right]+A,
\nonumber \\
&&A\equiv\frac{1}{8}\sum_{\mu\nu}
\left[|\psi_{\sigma}({\bf x+a^{\mu}+a^{\nu}})\rangle-|\psi_{\sigma}({\bf x+a^{\mu}-a^{\nu}})\rangle-|\psi_{\sigma}({\bf x-a^{\mu}+a^{\nu}})\rangle+|\psi_{\sigma}({\bf x-a^{\mu}-a^{\nu}})\rangle\right].
\label{psixa_lattice_expansion}
\end{eqnarray}
The overlap is then calculated by
\begin{eqnarray}
&&|\langle\psi_{\sigma}({\bf x})|\psi_{\sigma}({\bf x+a})\rangle|^{2}
\nonumber \\
&&=1+\frac{1}{2}\sum_{\mu}\left[\langle\psi_{\sigma}({\bf x})|\psi_{\sigma}({\bf x+a}^{\mu})\rangle-\langle\psi_{\sigma}({\bf x})|\psi_{\sigma}({\bf x-a}^{\mu})\rangle+\langle\psi_{\sigma}({\bf x+a}^{\mu})|\psi_{\sigma}({\bf x})\rangle-\langle\psi_{\sigma}({\bf x-a}^{\mu})|\psi_{\sigma}({\bf x})\rangle\right]
\nonumber \\
&&+\frac{1}{8}\langle\psi_{\sigma}({\bf x})|A+\frac{1}{8}A^{\dag}|\psi_{\sigma}({\bf x})\rangle
\nonumber \\
&&+\frac{1}{4}\sum_{\mu\nu}\left[\langle\psi_{\sigma}({\bf x+a}^{\mu})|-\langle\psi_{\sigma}({\bf x-a}^{\mu})|\right]|\psi_{\sigma}({\bf x})\rangle\langle\psi_{\sigma}({\bf x})|\left[|\psi_{\sigma}({\bf x+a}^{\nu})\rangle-|\psi_{\sigma}({\bf x-a}^{\nu})\rangle\right]+{\cal O}(a^{3})\equiv 1-g_{\mu\nu}a^{\mu}a^{\nu},
\label{psixpsixpa_square_expansion}
\end{eqnarray}
which defines the leading order correction to the overlap $g_{\mu\nu}a^{\mu}a^{\nu}$. We observe that if one demands $|\psi_{\sigma}({\bf x+a})\rangle$ to be also normalized up to second order in ${\bf a}^{\mu}$, then using Eq.~(\ref{psixa_lattice_expansion}) yields
\begin{eqnarray}
&&\langle\psi_{\sigma}({\bf x+a})|\psi_{\sigma}({\bf x+a})\rangle=1
\nonumber \\
&&+\frac{1}{2}\sum_{\mu}\left[\langle\psi_{\sigma}({\bf x})|\psi_{\sigma}({\bf x+a}^{\mu})\rangle-\langle\psi_{\sigma}({\bf x})|\psi_{\sigma}({\bf x-a}^{\mu})\rangle+\langle\psi_{\sigma}({\bf x+a}^{\mu})|\psi_{\sigma}({\bf x})\rangle-\langle\psi_{\sigma}({\bf x-a}^{\mu})|\psi_{\sigma}({\bf x})\rangle\right]
\nonumber \\
&&+\frac{1}{8}\langle\psi_{\sigma}({\bf x})|A+\frac{1}{8}A^{\dag}|\psi_{\sigma}({\bf x})\rangle
+\frac{1}{4}\sum_{\mu\nu}\left[\langle\psi_{\sigma}({\bf x+a}^{\mu})|-\langle\psi_{\sigma}({\bf x-a}^{\mu})|\right]\left[|\psi_{\sigma}({\bf x+a}^{\nu})\rangle-|\psi_{\sigma}({\bf x-a}^{\nu})\rangle\right]+{\cal O}(a^{3}),
\end{eqnarray}
and hence the second line of the order ${\cal O}(a)$ and the third line of the order ${\cal O}(a^{2})$ of this equation must vanish individually. Putting these observations back to Eq.~(\ref{psixpsixpa_square_expansion}), we obtain 
\begin{eqnarray}
&&g_{\mu\nu}a^{\mu}a^{\nu}=\frac{1}{4}\sum_{\mu\nu}\left[\langle\psi_{\sigma}({\bf x+a}^{\mu})|-\langle\psi_{\sigma}({\bf x-a}^{\mu})|\right]\left[|\psi_{\sigma}({\bf x+a}^{\nu})\rangle-|\psi_{\sigma}({\bf x-a}^{\nu})\rangle\right]
\nonumber \\
&&-\frac{1}{4}\sum_{\mu\nu}\left[\langle\psi_{\sigma}({\bf x+a}^{\mu})|-\langle\psi_{\sigma}({\bf x-a}^{\mu})|\right]|\psi_{\sigma}({\bf x})\rangle\langle\psi_{\sigma}({\bf x})|\left[|\psi_{\sigma}({\bf x+a}^{\nu})\rangle-|\psi_{\sigma}({\bf x-a}^{\nu})\rangle\right].
\label{gmunu_sum_lattice}
\end{eqnarray}
This leads to the formula for the quantum metric on the lattice 
\begin{eqnarray}
&&g_{\mu\nu}({\bf x},\sigma)=\frac{1}{8a^{\mu}a^{\nu}}\left[\langle\psi_{\sigma}({\bf x+a}^{\mu})|-\langle\psi_{\sigma}({\bf x-a}^{\mu})|\right]\left[|\psi_{\sigma}({\bf x+a}^{\nu})\rangle-|\psi_{\sigma}({\bf x-a}^{\nu})\rangle\right]
\nonumber \\
&&+\frac{1}{8a^{\mu}a^{\nu}}\left[\langle\psi_{\sigma}({\bf x+a}^{\nu})|-\langle\psi_{\sigma}({\bf x-a}^{\nu})|\right]\left[|\psi_{\sigma}({\bf x+a}^{\mu})\rangle-|\psi_{\sigma}({\bf x-a}^{\mu})\rangle\right]
\nonumber \\
&&-\frac{1}{4a^{\mu}a^{\nu}}\left[\langle\psi_{\sigma}({\bf x+a}^{\mu})|-\langle\psi_{\sigma}({\bf x-a}^{\mu})|\right]|\psi_{\sigma}({\bf x})\rangle\langle\psi_{\sigma}({\bf x})|\left[|\psi_{\sigma}({\bf x+a}^{\nu})\rangle-|\psi_{\sigma}({\bf x-a}^{\nu})\rangle\right]
\nonumber \\
&&=\frac{1}{8a^{\mu}a^{\nu}}\sum_{n}\left[a_{n\sigma}({\bf x+a}^{\mu})-a_{n\sigma}({\bf x-a}^{\mu})\right]\left[a_{n\sigma}^{\ast}({\bf x+a}^{\nu})-a_{n\sigma}^{\ast}({\bf x-a}^{\nu})\right]
\nonumber \\
&&+\frac{1}{8a^{\mu}a^{\nu}}\sum_{n}\left[a_{n\sigma}({\bf x+a}^{\nu})-a_{n\sigma}({\bf x-a}^{\nu})\right]\left[a_{n\sigma}^{\ast}({\bf x+a}^{\mu})-a_{n\sigma}^{\ast}({\bf x-a}^{\mu})\right]
\nonumber \\
&&-\frac{1}{4a^{\mu}a^{\nu}}\left\{\sum_{n}\left[a_{n\sigma}({\bf x+a}^{\mu})-a_{n\sigma}({\bf x-a}^{\mu})\right]a_{n\sigma}^{\ast}({\bf x})\right\}\left\{\sum_{n'}a_{n'\sigma}({\bf x})|\left[a_{n'\sigma}^{\ast}({\bf x+a}^{\nu})-a_{n'\sigma}^{\ast}({\bf x-a}^{\nu})\right]\right\},
\label{gmunu_lattice_expression}
\end{eqnarray}
\end{widetext}
that correctly sums to Eq.~(\ref{gmunu_sum_lattice}), is symmetric in the two indices $\mu$ and $\nu$, and recovers the usual derivative expression in the continuous limit $a_{\mu}\rightarrow 0$. Equation (\ref{gmunu_lattice_expression}) serves as a concrete numerical recipe to calculate the metric on discrete lattice points of a rectangular lattice, and is applicable to disordered systems since our derivation remains valid up to leading order in ${\bf a}^{\mu}$. The formula also suggests to implement the derivative $\partial_{\mu}$ on discrete lattice sites by the central difference, so we adopt the central difference in the calculation of all the differential geometrical properties given in Eq.~(\ref{Christoffel_Riemann_Ricci}), allowing them to be defined on discrete lattice sites.

The metric in Eq.~(\ref{gmunu_lattice_expression}) is invariant under the inversion of the increment $\delta{\bf x}={\bf a}^{\mu}\rightarrow\delta{\bf x}=-{\bf a}^{\mu}$, meaning that $|\langle\psi_{\sigma}({\bf x})|\psi_{\sigma}({\bf x+a})\rangle|$ and $|\langle\psi_{\sigma}({\bf x})|\psi_{\sigma}({\bf x-a})\rangle|$ give the same quantum distance under this central difference formalism. This is in line with the usual definition of distance in differential geometry $ds^{2}=g_{\mu\nu}\delta x^{\mu}\delta x^{\nu}$ in a small region around ${\bf x}$ that should be the same under $\delta{\bf x}={\bf a}^{\mu}\rightarrow\delta{\bf x}=-{\bf a}^{\mu}$ provided ${\bf a}^{\mu}$ is small enough. Moreover, the pattern of the metric is symmetric around the impurity site as shown in Fig.~\ref{fig:2Dmetal_Chern_figure}. Alternatively, one may use the forward or backward difference to expand the $|\psi_{\sigma}({\bf x+a})\rangle$ in Eq.~(\ref{psixa_lattice_expansion}). For instance, under the forward difference
\begin{eqnarray}
|\psi_{\sigma}({\bf x+a})\rangle&=&|\psi_{\sigma}({\bf x})\rangle
\nonumber \\
&+&\sum_{\mu}\left[|\psi_{\sigma}({\bf x+a}^{\mu})\rangle-|\psi_{\sigma}({\bf x})\rangle\right]+A',\;\;\;
\end{eqnarray}
and follow the same recipe above to construct the metric. However, the metric constructed this way changes under the inversion of the increment $\delta{\bf x}={\bf a}^{\mu}\rightarrow\delta{\bf x}=-{\bf a}^{\mu}$, which does not agree with the usual requirement in differential geometry, and moreover the spatial profile of the metric is not symmetric around the impurity site. Thus we disregard the forward and backward difference scenarios, and use only the central difference scenario presented above. 







\bibliography{Literatur_abbreviated}

\end{document}